\documentclass{PoS}
\usepackage{url}
\usepackage{amsmath,amsbsy}
\usepackage{multirow}

\newcommand{\Cyg}{\mbox{Cyg\,X-1}}
\newcommand{\Chandra}{\textsl{Chandra}}
\newcommand{\RXTE}{\textsl{RXTE}}
\newcommand{\XMM}{\textsl{XMM}}
\newcommand{\Suzaku}{\textsl{Suzaku}}
\newcommand{\Integral}{\textsl{INTEGRAL}}
\newcommand{\Swift}{\textsl{Swift}}
\newcommand{\AGILE}{\textsl{AGILE}}

\newcommand{\boldvec}[1]{\vec{\pmb{#1}}}
\newcommand{\vecr}{\boldvec{r}}
\newcommand{\vecv}{\boldvec{v}}
\newcommand{\veca}{\boldvec{a}}
\newcommand{\vecl}{\boldvec{l}}

\newcommand{\highlight}[1]{\emph{\textbf{#1}}}
\newcommand{\vskipAfterSection}{\vskip -0.25cm}

\title{Multi-Satellite Observations of Cygnus\,X-1\\\Large to Study the Focused Wind and Absorption Dips}
\ShortTitle{Multi-Satellite Observations of Cygnus\,X-1}

\author{\speaker{{\large Manfred Hanke}},\quad\quad
        J\"orn Wilms,\quad\quad
        Moritz B\"ock,\quad\quad
        \dots\footnote{%
         Felix F\"urst, Laura Barrag\'an, Ingo Kreykenbohm and Stefan Pirner
         from the Dr.~Karl Remeis-Observatory / ECAP have also contributed to this work.}\\
        Dr.~Karl Remeis-Observatory, Bamberg / Erlangen Centre for Astroparticle Physics, Germany\\
        E-mail: \email{Manfred.Hanke@sternwarte.uni-erlangen.de}}
\author{Michael A.~Nowak,\quad\quad
        Norbert S.~Schulz\\
        Massachusetts Institute of Technology, Kavli Institute for Astrophysics and Space Research, CXC}
\author{Katja Pottschmidt\\
        CRESST \& NASA Goddard Space Flight Center / CSST UMBC / CASS UCSD}
\author{Julia C.~Lee\\
        Harvard University, Department of Astronomy (Harvard-Smithsonian Center for Astrophysics)}

\abstract{
  High-mass X-ray binary systems are powered by the stellar wind of their donor stars.
  The X-ray state of Cygnus\,X-1 is correlated with the properties of the wind
  which defines the environment of mass accretion.
  \Chandra-HETGS observations close to orbital phase 0
  allow for an analysis of the photoionzed stellar wind at high resolution,
  but because of the strong variability due to soft X-ray absorption dips,
  simultaneous multi-satellite observations are required to track and understand the continuum, too.
  Besides an earlier joint \Chandra{} and \RXTE{} observation,
  we present first results from a recent campaign
  which represents the best broad-band spectrum of \Cyg{} ever achieved:
  On 2008 April 18/19 we observed this source with \textsl{XMM-Newton},
  \Chandra, \Suzaku, \RXTE, \Integral, \Swift, and \AGILE{} in X- and $\gamma$-rays,
  as well as with VLA in the radio.
  After~superior conjunction of the black hole, we detect soft X-ray absorption dips
  likely due to clumps in the focused wind covering $\ge95\,\%$ of the X-ray source,
  with column densities likely to be of several $10^{23}\,$cm$^{-2}$,
  which also affect photon energies above 20\,keV via Compton scattering.
}

\FullConference{VII Microquasar Workshop: Microquasars and Beyond\\September 1-5 2008\\Fo\c{c}a, Izmir, Turkey}

\begin{document}

%%%%%%%%%%%%%%%%%%%%%%
\section*{Introduction} \vskipAfterSection
%%%%%%%%%%%%%%%%%%%%%%
One of the currently still puzzling issues in our understanding of microquasars
is the existence of the different states (low/hard, high/soft, very high/steep power-law, \dots),
which distinguish themselves by the soft X-ray luminosity, the X-ray spectrum, the timing properties,
and also the nature of the jet and its radio spectrum.
One assumes that these states with their in many ways different properties
correspond to different configurations of the accretion flow
which cause different channels or efficiencies of energy dissipation.
\highlight{For a high-mass X-ray binary system,
the global framework of mass accretion onto the black hole
is defined by the powering wind of the donor star}.
One possibility is that changes in the wind properties (like density, velocity, ionization parameter, \dots)
might finally trigger state transitions.

This contribution focuses on the wind of the well-known persistently bright X-ray source \highlight{Cygnus\,X-1}.
Its companion star HDE\,226868, a $\sim$18$\,M_\odot$ O9.7 type supergiant \cite{Herrero1995},
fills 97\,\% of its Roche lobe \cite{GiesBolton1986_II}.
The stellar wind is therefore strongly focused due to tidal interactions \cite{FriendCastor1982,GiesBolton1986_III}.
The detailed structure of this wind
defines the environment of mass accretion onto the black hole on the largest scale.
At a luminosity of $L_{0.5-250\,\mathrm{keV}}=3\times10^{37}\,$erg\,s$^{-1}$
(which is obtained already in the low/hard state \cite{Hanke2008}),
the X-ray source produces a considerable feedback on the wind by photoionization \cite{Blondin1994},
which can create a complex wind structure within the binary system.
The parameters inferred from the H$\alpha$ emission line
are (anti-)correlated with the X-ray flux:
in the low/hard state, the wind is probably denser and therefore less strongly ionized,
which allows for a more efficient wind acceleration
and thus a lower accretion rate onto the black hole \cite{Gies2003}.

We have obtained multi-satellite observations of \Cyg{} in the hard state
close to superior conjunction of the black hole,
i.e., at phase $\phi_\mathrm{orb}\approx0$ in the 5.6\,day binary orbit \cite{Brocksopp1999,Gies2003},
which are ideal to investigate the wind:
at the low inclination $i\approx35^\circ$~\cite{Herrero1995},
no eclipses of \Cyg{} by its companion star can be observed.
Nevertheless, absorption dips, which are likely related to the focused stream covering the X-ray source,
are often detected close to $\phi_\mathrm{orb}=0$~\cite{BalucinskaChurch2000}.

%%%%%%%%%%%%%%%%%%%%%%%%%%%%%%%%%%%%%%%%%%%%%%%%%%%%%%%%%%%%%%
\section{The \Chandra{} and \RXTE{} observation of 2003 April} \vskipAfterSection
%%%%%%%%%%%%%%%%%%%%%%%%%%%%%%%%%%%%%%%%%%%%%%%%%%%%%%%%%%%%%%

\subsection{Advantages of a multi-satellite observation}
%%%%%%%%%%%%%%%%%%%%%%%%%%%%%%%%%%%%%%%%%%%%%%%%%%%%%%%%
\begin{figure}\centering
 \includegraphics[width=0.8\textwidth]{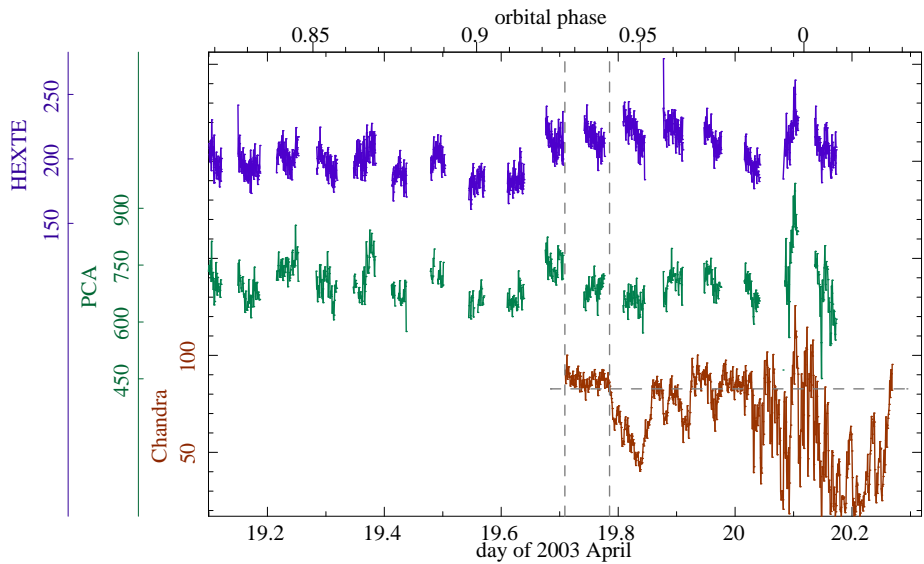}
 \caption{
  Light curves of \Cyg{} during the simultaneous \Chandra\,/\,\RXTE{} observation of 2003 April.
  Top:~\RXTE-HEXTE (20--250\,keV).
  Middle: \RXTE-PCA (4--20\,keV).
  Bottom: \Chandra-HETGS (\mbox{0.5--12\,keV} 1$^\mathrm{st}$ order events).
  The dashed lines define the non-dip times.
 }
 \label{fig:RXTElightcurve}
\end{figure}

In the first part of this contribution, we show some of the results
from a simultaneous observation of \Cyg{} performed by \Chandra{} and \RXTE{}
in 2003 April shortly before $\phi_\mathrm{orb}=0$.
(More details are given by Hanke et.~al \cite{Hanke2008,Hanke2009}.)
The light curves shown in Fig.~\ref{fig:RXTElightcurve}
clearly show the \highlight{advantages of a multi-satellite observation}:
Due to \RXTE's low Earth orbit, there are data gaps of $\sim$45\,minutes length
in each \RXTE-orbit (lasting $\sim$90\,minutes)
from passages through the South Atlantic Anomaly (SAA).
The high variability which is present on timescales of hours
and which is related to the stellar wind
could hardly be investigated with the \RXTE{} data alone.
The (continuous) soft X-ray light curve taken by \Chandra{}
is shaped by absorption dips.
Only the joint spectral analysis of the non-dip continuum
allows to combine the strength of the different instruments.
For example, the PCA spectrum clearly requires an Fe K$\alpha$ emission line at 6.4\,keV,
but the resolution is not good enough to decide
whether it is narrow or relativistically broadened.
In the simultaneous \Chandra{} spectrum,
a narrow component is detected (see Fig.~\ref{fig:spectrum}),
but its flux is not sufficient
to explain the PCA feature, from which we infer
the presence of a broad iron line as well.
(More details on iron lines in \Cyg{} are given by M.\,A.~Nowak
in his contribution to this volume \cite{Nowak2008_MQW7}.)

\subsection{The non-dip spectrum} \vskipAfterSection
%%%%%%%%%%%%%%%%%%%%%%%%%%%%%%%%%
The non-dip spectrum can be obtained by filtering for high count rates,
see the dashed horizontal line in Fig.~\ref{fig:RXTElightcurve}.
The results of this subsection are thoroughly described by Hanke et al.~(2008)~\cite{Hanke2008}.

The \Chandra-HETGS spectrum of the bright source \Cyg{}
shows \highlight{neutral absorption} signatures of the interstellar medium
(and probably also source environment).
We detect structured absorption edges, namely
the Fe L$_2$ and L$_3$ edges at 17.2 and 17.5\,\AA{} (0.721 and 0.708\,keV)
-- which J.\,C.~Lee describes in detail in her contribution in this volume --,
the O K-edge at 22.8\,\AA{} (0.544\,keV) with the K$\alpha$ resonance absorption line,
and the Ne K-edge at 14.3\,\AA{} (0.867\,keV) with the K$\beta$ resonance absorption line.
These complex absorption edges can be modelled with \texttt{tbnew}\footnote{~See~
 \url{http://pulsar.sternwarte.uni-erlangen.de/wilms/research/tbabs}.},
an improved version of \texttt{tbvarabs}, which contains those fine structures.

\begin{figure}
 \includegraphics[width=\textwidth]{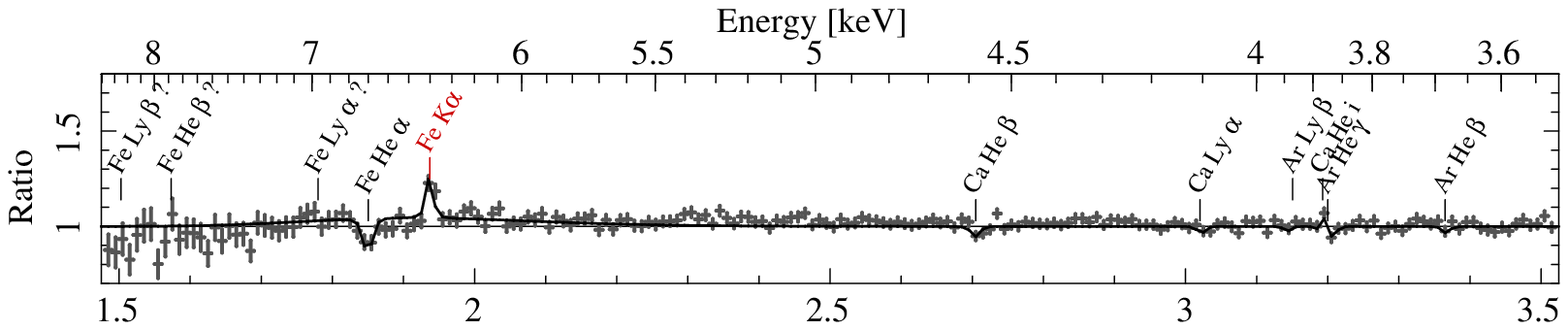}
 \includegraphics[width=\textwidth]{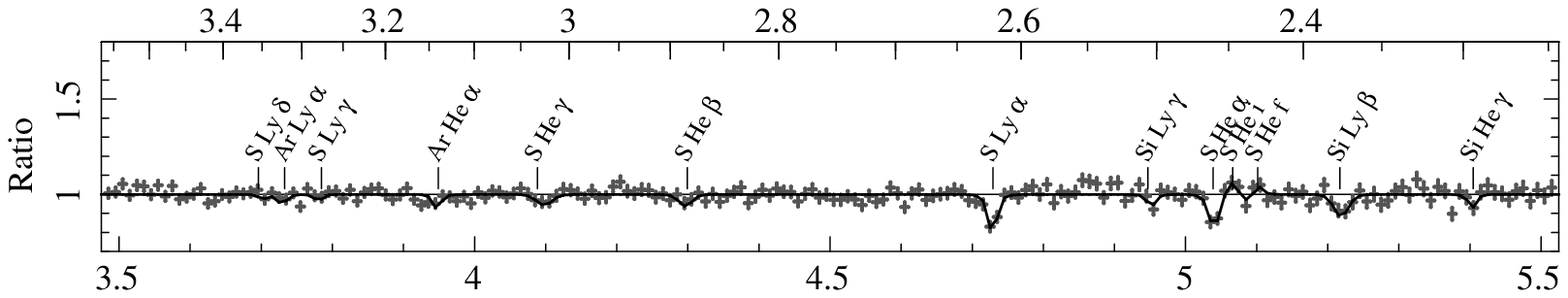}
 \includegraphics[width=\textwidth]{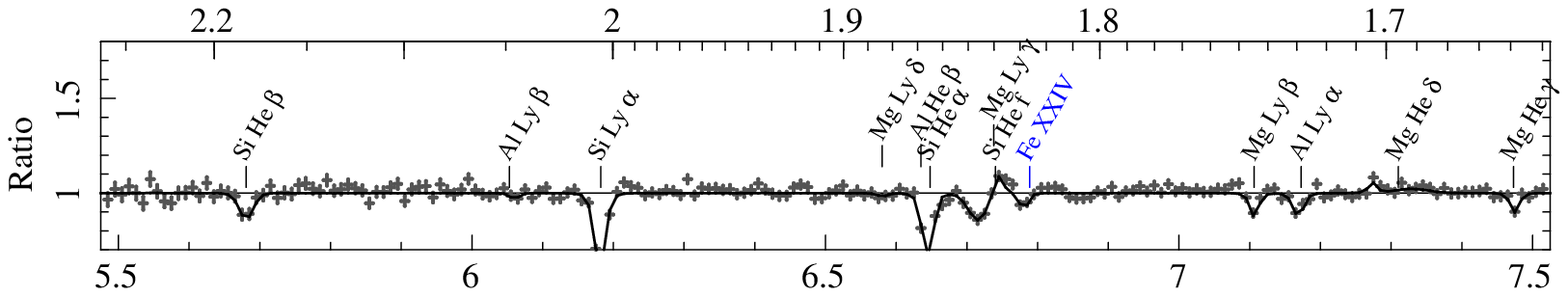}
 \includegraphics[width=\textwidth]{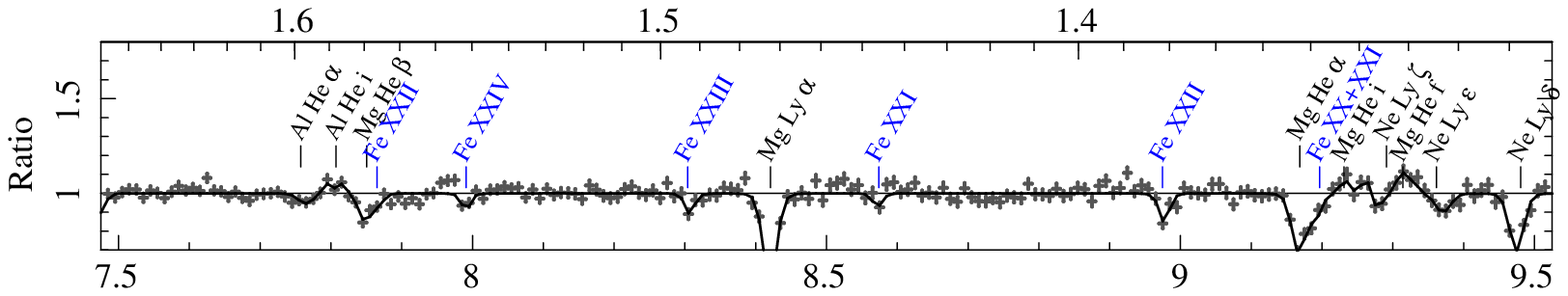}
 \includegraphics[width=\textwidth]{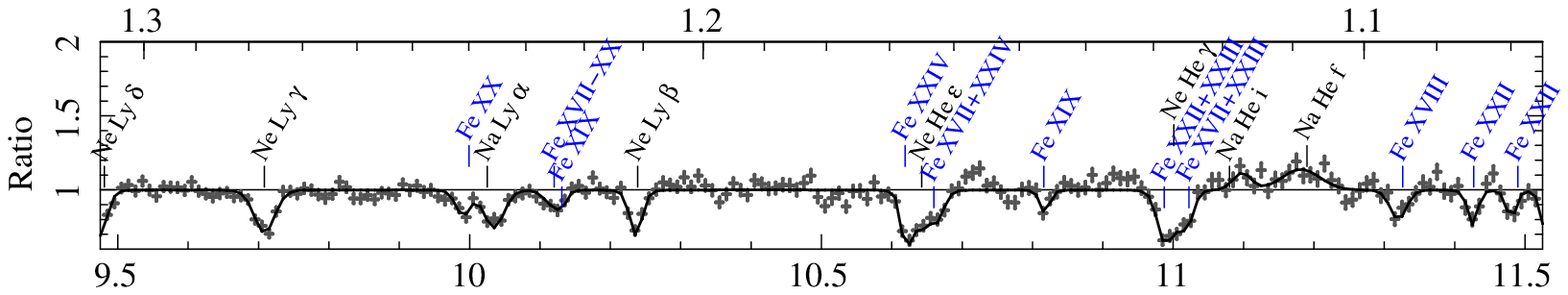}
 \includegraphics[width=\textwidth]{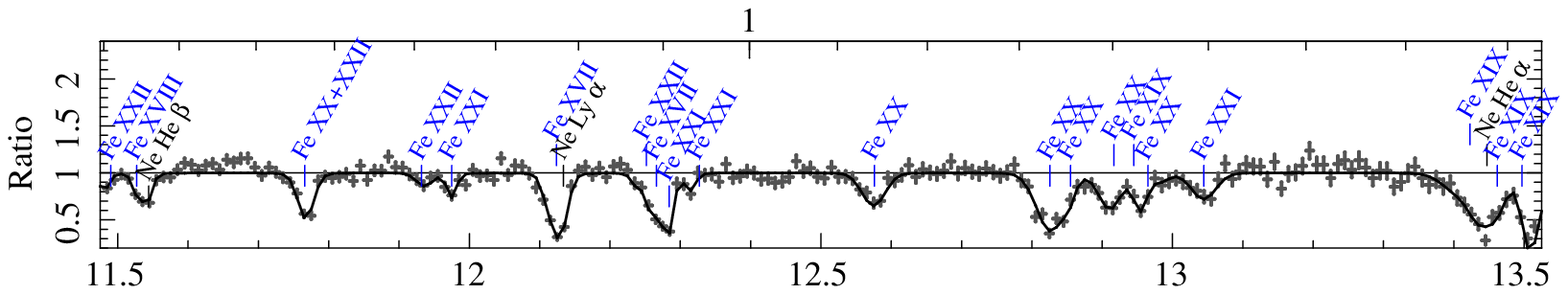}
 \includegraphics[width=\textwidth]{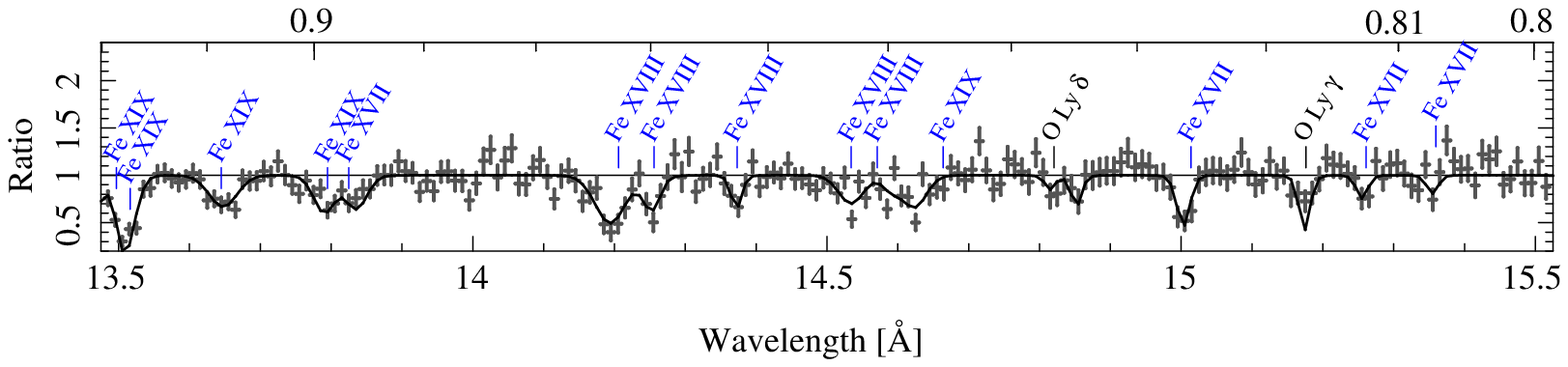}
 \caption{
  \Chandra-HETGS \Cyg{} non-dip spectrum 
  with identification and fits of absorption lines
  of \mbox{H- and He-like} ions (Lyman and Helium series; black labels)
  and Fe L-shell ions (blue labels).
  Note the narrow and broad Fe K$\alpha$ line (red) at 1.94\,\AA{}
  ($\equiv$6.4\,keV, see second $x$- (energy-) axis on top of each panel).
 }
 \label{fig:spectrum}
\end{figure}

The non-dip spectrum is dominated by \highlight{absorption lines from the stellar wind}
of highly ionized ions, namely H- and He-like O, Ne, Na, Mg, Al, Si, S, Ar, Ca, and Fe,
as well as many Fe L-shell absorption lines, see Fig.~\ref{fig:spectrum}.
We fit these features with Gaussian absorption lines and infer equivalent widths and
Doppler shifts with respect to the rest wavelengths of the identified transitions.
In a second approach, we have defined a fit-function
(in \textsc{ISIS} \cite{Houck2000,NobleNowak2008})
for the whole series of absorption lines of an ion.
Theoretical Voigt profiles were used which take into account
the oscillator strengths of the strongest transitions.
The ion's column density is a direct fit-parameter of the model,
avoiding an explicit curve of growth analysis.
The \highlight{velocity shifts} inferred from both methods
are \highlight{rather low}, mostly $|v|<200\,$km\,s$^{-1}$.
\begin{figure}\centering
 \begin{minipage}{0.5\textwidth}
  \includegraphics[width=\columnwidth,viewport=1 90 455 455, clip]{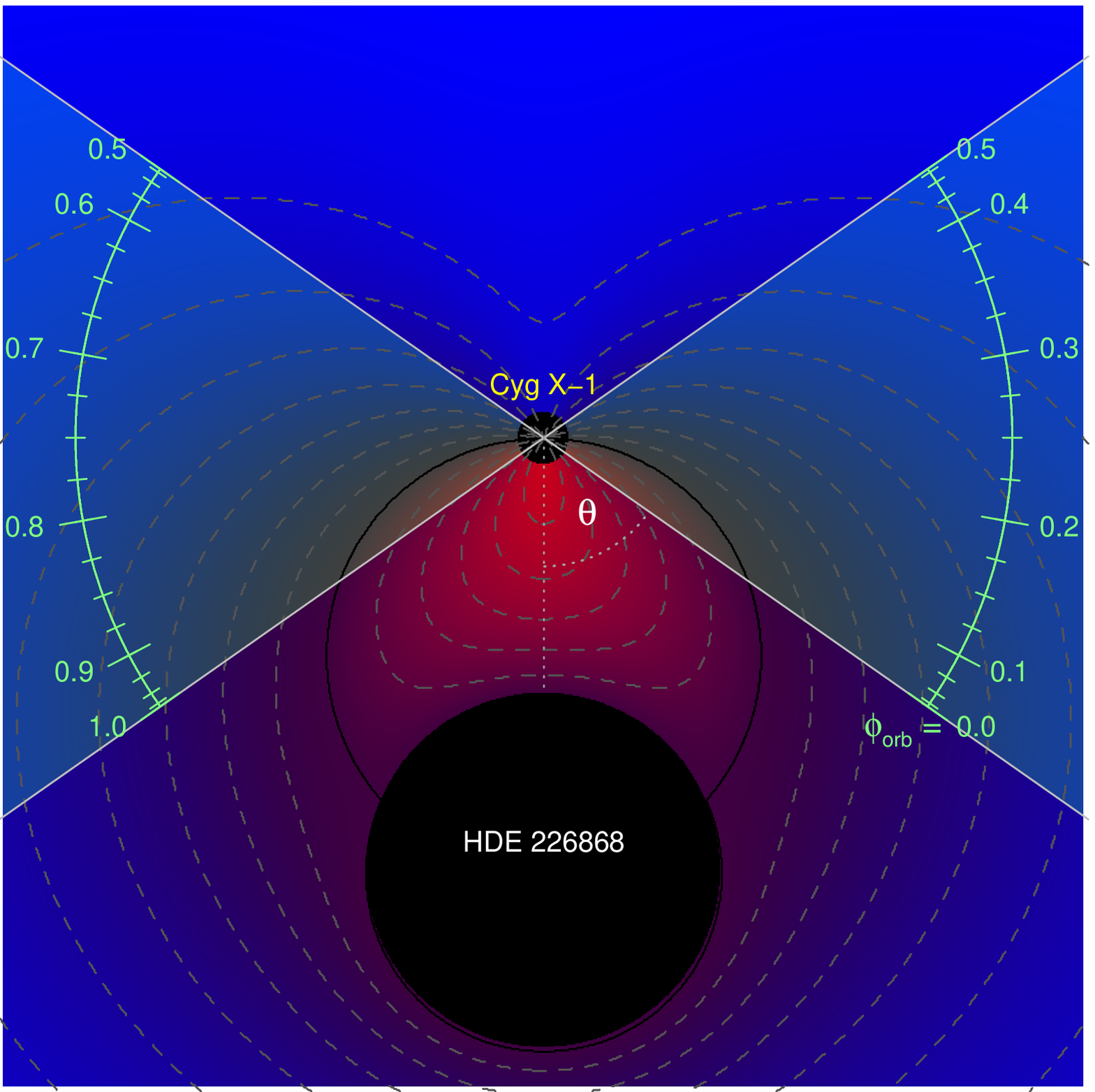}
 \end{minipage}\hfill
 \begin{minipage}{0.45\textwidth}
 \caption[]{
  Wind velocity, projected against the black hole.
  Absorption lines with low velocity $v_\mathrm{rad}$ can be observed from the dark purple region
  close to the black Thales circle of $v_\mathrm{rad}=0$,
  $v_\mathrm{rad}>0$ can only be observed from the red region inside this circle,
  and $v_\mathrm{rad}<0$ only from the blue region outside of it.
  The brighter the red or blue, the higher is $|v_\mathrm{rad}(\vecr)|$.
  The dashed gray lines are contours of constant $v_\mathrm{rad}$ in steps of 200\,km\,s$^{-1}$.
  The highlighted area with the $\phi_\mathrm{orb}$ scale
  shows the region in the orbital plane with $|\cos\vartheta|\le\sin i$,
  which equivalently corresponds to all possible lines of sight (for $i=35^\circ$).
  \label{fig:proj_velocity}
 }
 \end{minipage}
\end{figure}
The latter result can be understood within a simple model for the \textbf{spherical wind},
e.g., a generalized CAK-model \cite{CastorAbbottKlein1975,LamersLeitherer1993}:\\
\begin{equation}
 \vecv(\vecr) \;\;=\;\; v(r) \cdot \frac{\vecr}{r}\;\quad\quad\mbox{with}\quad
 v(r) \;\;=\;\; v_0 \;\;+\;\; (v_\infty-v_0) \cdot \left(1-\frac{R_\star}{r}\right)^\beta 
 \label{eq:vr}
\end{equation}
For the velocity observed in X-ray absorption lines, one has to consider
the \highlight{projection against the black hole}, $v_\mathrm{rad}(\vecr) \:=\: \cos\alpha(\vecr) \cdot v(\vecr)\:$,
where $\alpha(\vecr)$ is the angle between wind velocity $\vecv(\vecr)$ and
the line of sight towards the black hole.
Figure~\ref{fig:proj_velocity} shows $v_\mathrm{rad}(\vecr)$ as a colorscale plot.
On the circle passing through the center of the star
and the black hole, $\alpha=90^\circ$ by Thales' theorem,
and thus $v_\mathrm{rad}=0$ -- independent of the assumed velocity law,
as long as it is spherically symmetric.\linebreak
The circle separates (possible) absorbing regions of the wind seen at a redshift (inside)
from those seen at a blueshift (outside).
The location of absorbers seen at a low velocity is thus well constrained.
It is important to realize that this result does not depend on the model for the wind velocity,
provided that $\vecv(\vecr)$ points radially away from the star.

Due to the system's inclination, not every line of sight in the orbital plane can actually be observed:
the angle $\vartheta$ between the inclined line of sight $\vecl = (\sin i,\;\: 0,\;\: \cos i)$
and the binary axis $\veca = \big(\cos(2\pi\phi_\mathrm{orb}),\; \sin(2\pi\phi_\mathrm{orb}),\; 0\big)$
is given by $\cos\vartheta = \vecl \,\circ\, \veca = \sin i\cdot\cos(2\pi\phi_\mathrm{orb})$.
The range of possible $\vartheta$ values is obviously limited by a low inclination.
(For a face-on geometry with $i=0^\circ$, there is clearly no orbital variation, i.e. $\vartheta\equiv90^\circ$,
 while every $0\le\vartheta\le180^\circ$ is possible for edge-on systems with $i=90^\circ$.)
The region with $|\cos\vartheta|\le\sin i$ is highlighted in Fig.~\ref{fig:proj_velocity}
between the crossing solid gray lines. For a spherically symmetrical wind,
this region in the orbital plane is equivalent to all lines of sight experienced by an inclined observer.
The green labels show the position of these lines of sight for given orbital phases $\phi_\mathrm{orb}$,
where $\phi_\mathrm{orb}$ and $1-\phi_\mathrm{orb}$ are identical.

As a result, the observation of low Doppler velocities in absorption lines
in front of the \mbox{X-ray} source constrains the location of the absorber
-- rather independently of the wind model.
This location is in good agreement with \highlight{\textsc{xstar} simulations of the photoionization zone}.
The number density of H-atoms, $n_\mathrm{H}$, is obtained from the velocity profile (Eq.~\ref{eq:vr})
and the continuity equation,
$ \dot M_\star \;=\; \mu\,m_\mathrm{H} \cdot n_\mathrm{H}(r) \cdot 4\pi r^2 \cdot v(r) $.
Here, $\mu=1.4$ is the mean molecular weight per H-atom and $m_\mathrm{H}$~is~the mass of an H-atom.
Inserting the numbers $v_\infty=2100\,\mathrm{km\,s}^{-1}$ and $R_\star=17\,R_\odot$ \cite{Herrero1995} gives:
\begin{equation}
 n_\mathrm{H}(r) \;=\; 2.2\!\times\!10^{10}\,\mathrm{cm}^{-3} \cdot d(r/R_\star) \;\;,
\;\;\mbox{where}\;\;
 d(x) \;=\; \frac{x^{-2}}{v_0/v_\infty \,+\, \left(1-v_0/v_\infty \right)\cdot\left(1-1/x\right)^\beta}
\end{equation}
Between the stellar surface and the black hole, $x=r/R_\star$ varies between 1 and 2.4
and the average wind density is $n_\mathrm{H}\approx10^{11}\,$cm$^{-3}$
for $v_0/v_\infty=v_\mathrm{therm}/v_\infty=0.01$ and a reasonable $\beta\approx1$.

\begin{figure}\centering
 \begin{minipage}{0.67\textwidth}
  \includegraphics[width=0.95\columnwidth]{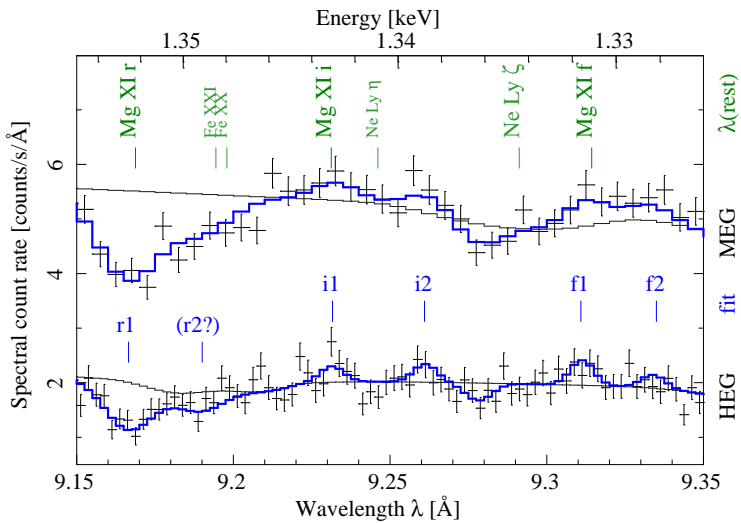}
 \end{minipage}
 \begin{minipage}{0.3\textwidth}
  \caption{
   The Mg\,\textsc{xi} triplet of resonance (r) absorption and
   intercombination (i) as well as forbidden (f) emission lines
   is detected with two pairs of \mbox{i- and} f-lines in the non-dip
   \Chandra-HETGS spectra of \Cyg.
   The red-shifted pair stems from an emitting plasma component of higher density
   and can be identified with the focused wind.
   \label{fig:MgTriplet}
  }
 \end{minipage}
\end{figure}

While these previous findings from the non-dip spectrum could be explained by a spherical wind only,
there are also signatures of a \highlight{focused wind}:
Figure~\ref{fig:MgTriplet} shows the Mg\,\textsc{xi} triplet of
$\rm 1s^2\,(^1S_0)\rightarrow1s2p\,(^1P_1)$ ``resonance'' (r),
$\rm 1s^2\,(^1S_0)\leftarrow1s2p\,(^3P_{1,2})$ ``intercombination'' (i)
and $\rm 1s^2\,(^1S_0)\leftarrow1s2s\,(^3S_1)$ ``forbidden'' (f) transitions.
The i- and f-emission lines show up as two components each:
one pair is almost at rest -- in accord with the r absorption line --,
while the other pair is seen at a strong redshift (500--900\,km\,s$^{-1}$).
The ratio $R=F($f$)/F($i$)$ of the line fluxes is usually used for density diagnostics \cite{PorquetDubau2000}.
However, in the presence of strong UV-radiation fields (such as in the vicinity of HDE\,226868),
the $R$ ratio is systematically decreased due to photodepletion of the $\rm 1s2s\,(^3S_1)$ level \cite{Kahn2001}.
Therefore, an absolute density analysis would overestimate the density.
One can, however, still infer the relative statement that the red-shifted pair of i- and f-lines
(with the lower $R$-ratio) stems from a much denser emitting plasma component.
Both its velocity and its density indicate that this component
can be identified with the focused wind.

\subsection{The dip-spectrum} 
\begin{figure}\centering
 \includegraphics[width=0.475\textwidth]{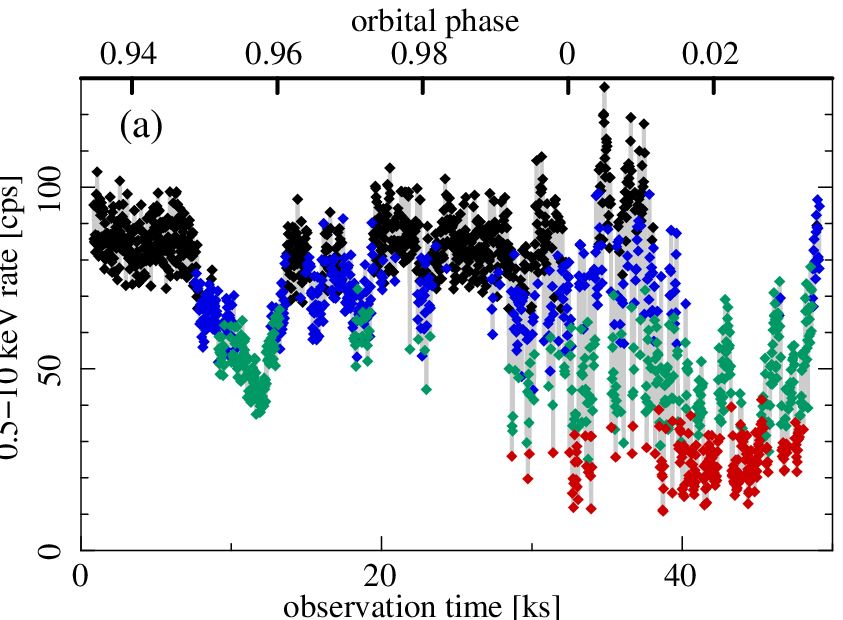}\hfill
 \includegraphics[width=0.475\textwidth]{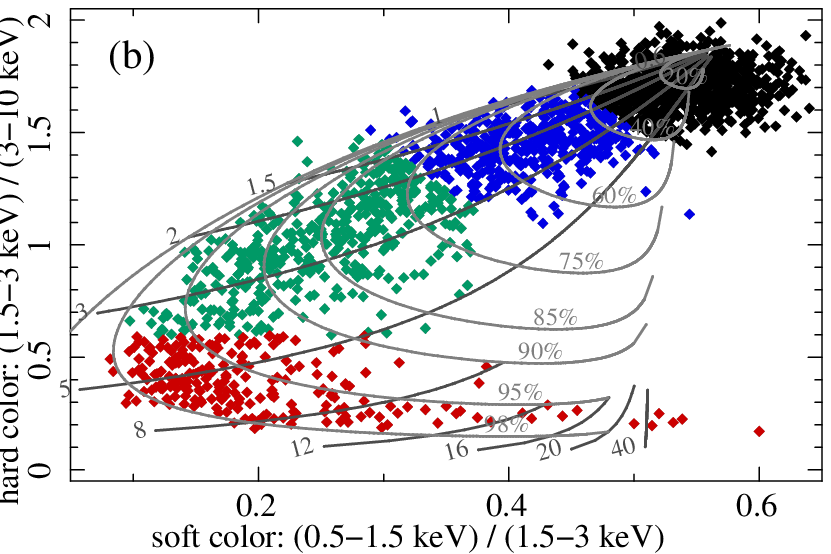}
 \caption[]{
   (a): \Chandra-HETGS (0.5--10\,keV) light curve (see also Fig.~\ref{fig:RXTElightcurve}).\quad
   (b): Color-color diagram from (0.5--1.5\,keV)$:$(1.5--3\,keV) and (1.5--3\,keV)$:$(3--10\,keV)
        count rate ratios in 25\,s intervals, explained by a grid of partial covering models
        (Eq.~1.3), assuming a power-law continuum ($\Gamma=1.73$).
        The gray labels indicate the covering fraction $f_\mathrm{c}$ in \% and
        the column density $N_\mathrm{H,2}$ of the additional absorber in $10^{22}\,$cm$^{-2}$.
 }
 \label{fig:colorcolor}
\end{figure}

\vskipAfterSection
%%%%%%%%%%%%%%%%%%%%%%%%%%%%%
During the dips (Fig.~\ref{fig:colorcolor}a),
the soft X-ray spectrum is considerably absorbed.
Defining the soft X-ray color as the ratio of the 0.5--1.5\,keV and the 1.5--3\,keV \Chandra{} count rate,
and the hard color as the ratio of 1.5--3\,keV and  3--10\,keV rate,
a color-color diagram (Fig.~\ref{fig:colorcolor}b) shows 
that neutral photoabsorption alone cannot explain the data.
The nose-like shape of the data points
can be explained by a \highlight{partial covering model}
for the photoelectric absorption:
\begin{equation}
 (1-  f_\mathrm{c})\cdot \mathrm{e}^{-N_\mathrm{H,1}^\mathrm{ISM}\cdot\sigma_\mathrm{bf}(E)}
 \;+\;f_\mathrm{c} \cdot \mathrm{e}^{-N_\mathrm{H,2}\cdot\sigma_\mathrm{bf}(E)}
 \label{eq:partialcov}
\end{equation}
We use a constant $N_\mathrm{H,1}^\mathrm{ISM}=5.4\!\times\!10^{21}\,\mathrm{cm}^{-2}$
and a power-law continuum with photon index \mbox{$\Gamma=1.73$}
as found with \RXTE-PCA for the non-dip spectrum \cite{Hanke2008}.
In this model, the strongest dips can be described by
an additional absorber with \mbox{$N_\mathrm{H,2}>4\!\times\!10^{23}\,\mathrm{cm}^{-2}$}
covering a fraction $f_\mathrm{c}$ between 95\,\% and 98\,\% of the X-ray source.

\begin{figure}
 \includegraphics[width=\textwidth]{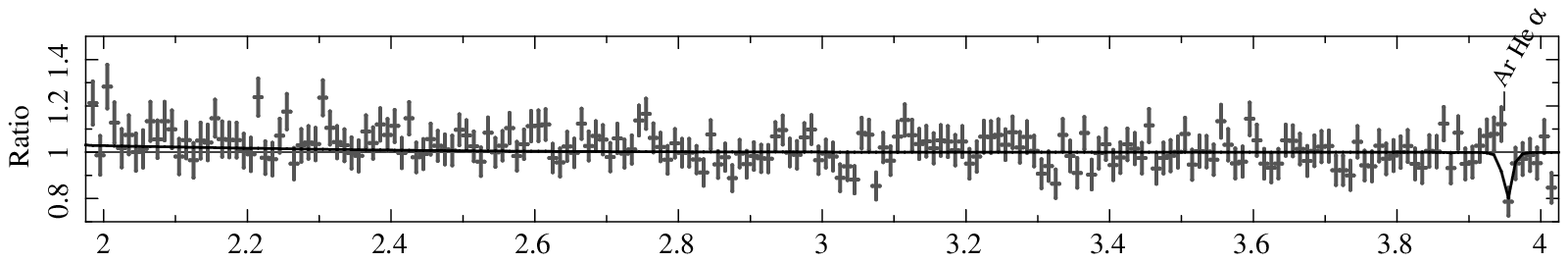}
 \includegraphics[width=\textwidth]{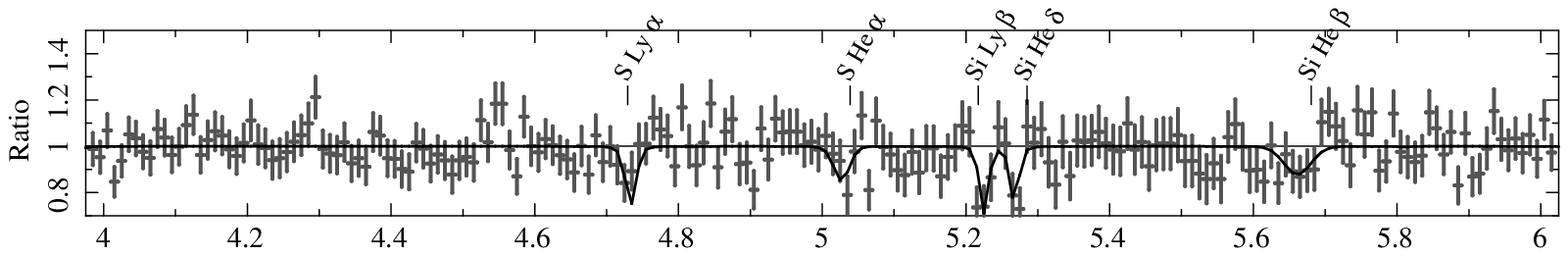}
 \includegraphics[width=\textwidth]{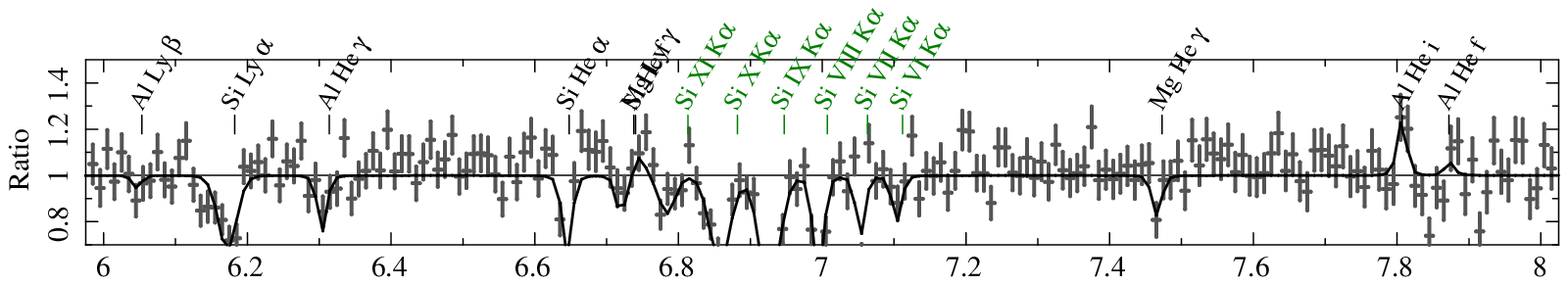}
 \includegraphics[width=\textwidth]{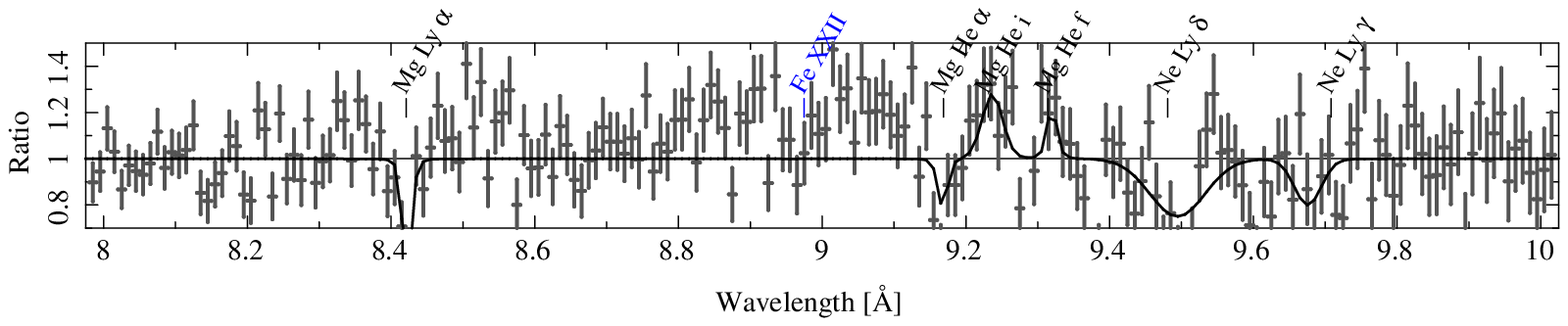}
 \caption[]{
  \Chandra-HETGS \Cyg{} spectrum of the first dip ($\phi_\mathrm{orb}<0.96$, see Fig.~\ref{fig:colorcolor}a).
  While the non-dip spectrum (Fig.~\ref{fig:spectrum}) showed silicon resonance K$\alpha$ absorption lines
  of only Si\,\textsc{xiv} and \textsc{xiii}
  (Si Ly$\alpha$ at 6.18\,\AA{} and Si He$\alpha$ at 6.65\,\AA),
  the dip spectrum also presents the lower ionization stages Si\,\textsc{vi}--\textsc{xi}
  (green labels).
 }
 \label{fig:dipspectrum}
\end{figure}

The high-resolution dip spectra (see, e.g., Fig.~\ref{fig:dipspectrum})
show strong absorption lines of \highlight{lower ionized ions}, e.g.,
resonance K$\alpha$ absorption lines of Si\,\textsc{vi}--\textsc{xi} between 6.8 and 7.1\,\AA,
which are not detected in the non-dip spectrum at all.
This result is consistent with an origin of the dips in colder matter in the focused wind
which covers the X-ray source.
(More details will be presented in \cite{Hanke2009}.)

%%%%%%%%%%%%%%%%%%%%%%%%%%%%%%%%%%%%%%%%%%%%%%%%%%%%
\section{The multi-satellite campaign of 2008 April} \vskipAfterSection
%%%%%%%%%%%%%%%%%%%%%%%%%%%%%%%%%%%%%%%%%%%%%%%%%%%%
For a \highlight{holistic study} of \Cyg{} and the focused wind of HDE\,226868,
further multi-satellite observations had to be performed.
The strong variability due to dipping requires to observe \Cyg{}
strictly simultaneously, if one wants to benefit from several different instruments
in order to study the wind and the absorption dips.
We originally proposed for a joint \XMM+\Chandra\footnote{~\mbox{Note that 
 scheduling an simultaneous observation with orbital phase constraints is not easy
 for these two satellites...}}
observation -- with \XMM's EPIC-pn camera for a high sensitivity for the spectral continuum shape
through the broad iron line region, \Chandra's HETGS for high-resolution wind spectroscopy,
and \XMM's RGS for a high-resolution spectrum at energies below the O-edge.
\RXTE{} was available to provide the broadband continuum spectrum up to 250\,keV.
We were also able to obtain simultaneous \Integral{} time
-- extending the spectral coverage up to $\sim$2\,MeV --, as well as \textsl{Suzaku}
-- allowing for an independent confirmation of the spectral continuum,
see Table~\ref{tab:log}.
We were furthermore awarded a few~ks of simultaneous \Swift{} TOO time.
\AGILE{} did not detect the source in the 30\,MeV--50\,GeV band.
\Cyg{} has also been monitored in the radio by the \textsl{VLA} during this observation.

\begin{table}{\centering
 \caption{Multi-satellite observation of Cygnus\,X-1 on 2008 April 18/19.}
 \label{tab:log}
 \begin{tabular}{cccl}
  \hline
  \hline
  Satellite & Start & Stop & Notes\\
  \hline
  \XMM      & 2008-04-18, 12:07:27 & 2008-04-19, 04:25:46 \\
  \Integral & 2008-04-18, 12:35:32 & 2008-04-19, 23:00:45 & $\subseteq$ \{obs. of the Cyg-region\}\\
  \multirow{2}{*}{\raisebox{0.05in}{\RXTE}~~\Huge\{}
            & 2008-04-18, 13:21:36 & 2008-04-19, 04:22:39 & \multirow{2}{*}{{\Huge\}}~~\raisebox{0.05in}{with SAA-interruptions}} \\
            & 2008-04-19, 17:37:20 & 2008-04-19, 20:04:47 & \\
  \Suzaku   & 2008-04-18, 16:22:33 & 2008-04-19, 10:08:08 & with SAA-interruptions\\
  \multirow{2}{*}{\raisebox{0.05in}{\Chandra}~~\Huge\{}
            & 2008-04-18, 18:27:19 & 2008-04-19, 02:48:58 \\
            & 2008-04-19, 14:57:27 & 2008-04-19, 20:17:26 \\
  \hline
 \end{tabular}\\
 \ \\}
 Note that, according to \cite{Gies2003}, $\;\;\phi_\mathrm{orb}($2008-04-18, 15:53$)=0\;\;$
 and $\;\;\phi_\mathrm{orb}($2008-04-19, 20:00$)=0.20\;$.
\end{table}

As expected, the observation was again dominated by \highlight{dipping events}.
Most of them are restricted to the soft X-ray band (\Chandra, \XMM),
but a few are also detected above 4\,keV by the PCA on \RXTE{}
or even above 20\,keV by IBIS on \Integral.
A color-color diagram reveals the same nose-like shape as was found for the dips in 2003
(Fig.~\ref{fig:colorcolor}b), indicating the same possible origin of the dips
in partial covering of the X-ray source \cite{Nowak2008_MQW7,Hanke2009}.

\begin{figure}\centering
 \includegraphics[width=\textwidth]{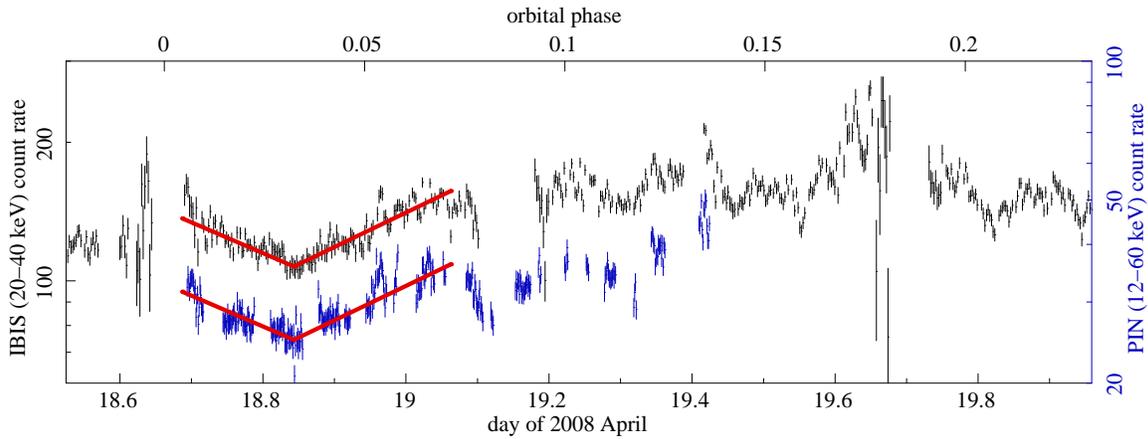}
 \caption{Light curves obtained with \Integral-IBIS in the 20--40\,keV band (top black data, left $y$ axis)
          and \Suzaku-PIN in the 12--60\,keV band (lower blue data, right $y$-axis)
	  with a time resolution of 200 and 96\,s,
          revealing an X-ray scattering trough shortly after superior conjunction,
          which lasts for 8.5 hours. During this time, the light curves can be described by
          two continuously jointed linear functions (red lines).
         }
 \label{fig:scatteringTrough}
\end{figure}

Due to the fast decrease of the photoelectric absorption cross section $\sigma_\mathrm{bf}$
above the ionization threshold,
X-ray photons with energies well above the iron edge at 7.1\,keV
can hardly be absorbed by the wind.
They can, however, still be scattered out of the line of sight.
Figure~\ref{fig:scatteringTrough} shows such an X-ray \highlight{scattering trough}
lasting more than 8 hours:
Between April~18,~$\sim$16:45 and April~19,~$\sim$1:15,
the 20--40\,keV count rate measured with \Integral-IBIS,
as well as the 12--60\,keV count rate measured with \Suzaku-PIN,
decrease until April~18,~20:15 and rise again afterwards.
A continuous model joining two linear functions
(a descending one followed by an ascending one)
can describe both light curves simultaneously
(with a scaling factor of 0.23 between the PIN and IBIS count rates)
remarkably well ($\chi^2_\mathrm{red}=2$).
If the 30\,\% reduction between the highest and the lowest count rate of this fit
(corresponding to non-dip and the strongest dip)
is due to Compton scattering, the optical depth $\tau=0.36$ corresponds to
$N_\mathrm{H}=6\!\times\!10^{23}$cm$^{-2}$.

\begin{figure}\centering
 \includegraphics[width=0.96\textwidth]{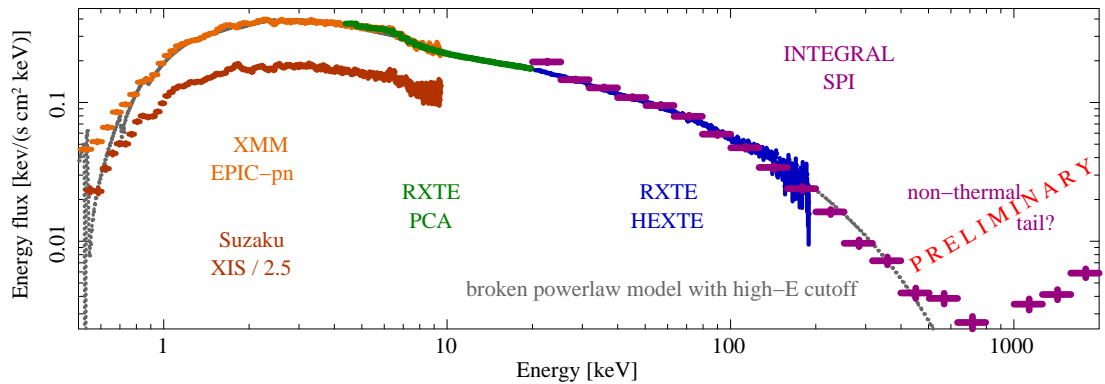}
 \caption{The 0.5\,keV--2\,MeV broadband spectrum of \Cyg,
          observed simultaneously with \XMM, \Suzaku, \RXTE{} and \Integral{}
          in 2008 April. The data below 300\,keV can be described consistently~by a broken power-law
          with exponential cutoff at high energies ($E_\mathrm{fold}\approx170\,$keV).
          ISIS'$\!$ (model independent) flux-corrected spectra (relying on the response matrices)
          seem to indicate a non-thermal tail in the SPI data.
         }
 \label{fig:broadbandspectrum}
\end{figure}

The \highlight{broadband spectrum} between 0.5\,keV and 2\,MeV,
assembled from various instruments (Fig.~\ref{fig:broadbandspectrum}),
reveals that \Cyg{} was again observed in the hard state.
Below 300\,keV, it can be described as a simple elbow-shaped broken power-law with high energy cutoff.
\RXTE-PCA's photon indices below and above $E_\mathrm{break}=9.8\,$keV,
namely $\Gamma_1=1.7$ and $\Gamma_2=1.4$, are amongst the hardest ones
found in the long term monitoring of \Cyg{} from 1999 to 2004 \cite{Nowak2008_MQW7,Wilms2006}.
Preliminary \Integral-SPI data seem to indicate a non-thermal tail in the spectrum above 400\,keV,
but further analysis through joint broadband fits with Comptonization models is still ongoing.

%%%%%%%%%%%%%%%%
\acknowledgments \vskipAfterSection
%%%%%%%%%%%%%%%%
We thank the organizing committees for this wonderful conference in Fo\c{c}a\,/\,Izmir, Turkey!
This~work was funded by the \textsl{Bundesministerium f\"ur Wirtschaft und Technologie}
through the \textsl{Deutsches Zentrum f\"ur Luft- und Raumfahrt} under contract 50OR0701.

%%%%%%%%%%%%%%%%%%%%%%%


\begin{thebibliography}{99}
%%%%%%%%%%%%%%%%%%%%%%%
\bibitem{BalucinskaChurch2000}
M.~{Ba{\l}uci{\'n}ska-Church}, M.~J. Church, P.~A. Charles, F.~Nagase,
  J.~LaSala, and R.~Barnard, \emph{The distribution of X-ray dips with orbital
  phase in Cygnus X-1}, MNRAS\, \textbf{311}, 861--868 (2000).

\bibitem{Blondin1994}
J.~M. Blondin, \emph{The shadow wind in high-mass X-ray binaries}, ApJ\,
  \textbf{435}, 756--766 (1994).

\bibitem{Brocksopp1999}
C.~Brocksopp, A.~E. Tarasov, V.~M. Lyuty, and P.~Roche, \emph{An improved
  orbital ephemeris for Cygnus~X-1}, A\&A\, \textbf{343}, 861--864 (1999).

\bibitem{CastorAbbottKlein1975}
J.~I. Castor, D.~C. Abbott, and R.~I. Klein, \emph{Radiation-driven winds in Of
  stars}, ApJ\, \textbf{195}, 157--174 (1975).

\bibitem{FriendCastor1982}
D.~B. Friend and J.~I. Castor, \emph{Radiation-driven winds in X-ray binaries},
  ApJ\, \textbf{261}, 293--300 (1982).

\bibitem{GiesBolton1986_II}
D.~R. Gies and C.~T. Bolton, \emph{The optical spectrum of HDE 226868 = Cygnus
  X-1. II.~Spectrophotometry and mass estimates}, ApJ\, \textbf{304}, 371--388
  (1986).

\bibitem{GiesBolton1986_III}
D.~R. Gies and C.~T. Bolton, \emph{The optical spectrum of HDE 226868 = Cygnus
  X-1. III. A focused stellar wind model for He II $\lambda$ 4686 emission}, ApJ\,
  \textbf{304}, 389--394 (1986).

\bibitem{Gies2003}
D.~R. Gies, C.~T. Bolton, J.~R. Thomson, W.~Huang, M.~V. McSwain, R.~L. Riddle,
  Z.~Wang, P.~J. Wiita, D.~W. Wingert, B.~{Cs{\'a}k}, and L.~L. Kiss,
  \emph{Wind accretion and state transitions in Cygnus X-1}, ApJ\,
  \textbf{583}, 424--436 (2003).

\bibitem{Hanke2008}
M.~Hanke, J.~Wilms, M.~A. Nowak, K.~Pottschmidt, N.~S. Schulz, and J.~C. Lee,
  \emph{X-ray spectroscopy of the focused wind in the Cygnus X-1 system with
  Chandra. I.~The non-dip spectrum in the low/hard state}, ApJ (2008),
  in press (ArXiv:0808.3771).

\bibitem{Hanke2009}
M.~Hanke, J.~Wilms, M.~A. Nowak, K.~Pottschmidt, N.~S. Schulz, and J.~C. Lee,
  \emph{X-ray spectroscopy of the focused wind in the Cygnus x-1 system with
  Chandra II.~The dip spectrum}, ApJ (2009), in
  preparation.

\bibitem{Herrero1995}
A.~Herrero, R.~P. Kudritzki, R.~Gabler, J.~M. Vilchez, and A.~Gabler,
  \emph{Fundamental parameters of galactic luminous OB stars. II.~A
  spectroscopic analysis of HDE 226868 and the mass of Cygnus X-1}, A\&A\,
  \textbf{297}, 556 (1995).

\bibitem{Houck2000}
J.~C. Houck and L.~A. Denicola, \emph{ISIS: An interactive spectral
  interpretation system for high resolution X-ray spectroscopy} in proceedings
  of \emph{Astronomical Data Analysis Software and Systems IX,} (N.~Manset,
  C.~Veillet, and D.~Crabtree, eds.), ASP Conf. Ser., no. 216, 2000, p.~591.

\bibitem{Kahn2001}
S.~M. Kahn, M.~A. Leutenegger, J.~Cottam, G.~Rauw, J.-M. Vreux, A.~J.~F. {den
  Boggende}, R.~Mewe, and M.~{G{\"u}del}, \emph{High resolution X-ray
  spectroscopy of zeta Puppis with the XMM-newton Reflection Grating
  Spectrometer}, A\&A\, \textbf{365}, L312--L317 (2001).

\bibitem{LamersLeitherer1993}
H.~J.~G.~L.~M. Lamers and C.~Leitherer, \emph{What are the mass-loss rates of O
  stars?}, ApJ\, \textbf{412}, 771--791 (1993).

\bibitem{NobleNowak2008}
M.~S. Noble and M.~A. Nowak, \emph{Beyond XSPEC: Toward highly configurable
  astrophysical analysis}, PASP\, \textbf{120}, 821--837 (2008).

\bibitem{Nowak2008_MQW7}
M.~A. Nowak, \emph{{Suzaku Observations of Cyg X-1}} in proceedings of
  \emph{VII Microquasar Workshop: Microquasars and Beyond}, 2008,
  (ArXiv:0810.1519).

\bibitem{PorquetDubau2000}
D.~Porquet and J.~Dubau, \emph{X-ray photoionized plasma diagnostics with
  helium-like ions. Application to warm absorber-emitter in active galactic
  nuclei}, A\&AS\, \textbf{143}, 495--514 (2000).

\bibitem{Wilms2006}
J.~Wilms, M.~A. Nowak, K.~Pottschmidt, G.~G. Pooley, and S.~Fritz,
  \emph{Long term variability of Cyg~X-1. IV. Spectral evolution 1999-2004},
  A\&A\, \textbf{447}, 245--261 (2006).
\end{thebibliography}
\end{document}